\documentclass[pra,twocolumn,showpacs,floatfix]{revtex4}
\usepackage{graphicx}
\begin{document}

\title{Metastability of persistent currents in trapped gases of atoms}
\author{K. K\"arkk\"ainen$^1$, J. Christensson$^1$, G. Reinisch$^2$,
G. M. Kavoulakis$^1$, and S. M. Reimann$^1$}
\affiliation{$^1$Mathematical Physics, Lund Institute of Technology,
P.O. Box 118, SE-22100 Lund, Sweden
\\
$^2$D{\' e}partement Cassiop{\' e}e, Observatoire de la
C{\^o}te d'Azur, BP 4229, 06304 - Nice C{\' e}dex 4, France}
\date{\today}

\begin{abstract}

We examine the conditions that give rise to metastable,
persistent currents in a trapped Bose-Einstein condensate. A
necessary condition for the stability of persistent currents is
that the trapping potential is not a monotonically increasing
function of the distance from the trap center. Persistent
currents also require that the interatomic interactions are
sufficiently strong and repulsive. Finally, any off-center
vortex state is shown to be unstable, while a driven gas shows
hysteresis.

\end{abstract}
 \maketitle

\section{Introduction}

Vapors of ultracold atoms provide an ideal laboratory for
testing the superfluid properties of quantum systems, including
-- among many other things -- persistent currents. These
phenomena are not only interesting from a theoretical point of
view, but may also have very important technological
applications \cite{appl}.

As opposed to the traditional studies on liquid Helium that is
confined in a box, gases of atoms may be confined in potentials
of various form, such as harmonic, anharmonic, etc. In
addition, the coupling constant between the atoms is tunable,
and it is even possible to change its sign, allowing us to
study both repulsive, as well as attractive effective
interactions between the atoms.

As we demonstrate below, the two above features allow one to
tune the stability of persistent currents in atomic systems by
appropriately choosing the functional form of the trapping
potential, the size/strength of the scattering length that
describes the low-energy two-body collisions between the atoms,
or the atom density/atom number.

Numerous theoretical and experimental studies have examined the
rotational properties of clouds of trapped atoms, the vortex
nucleation and stability, and more generally the superfluid
properties of these gases. For a review of the extensive work
on all these problems, we refer to the two articles by Leggett,
on superfluidity in general \cite{Leggett1}, and in trapped
gases \cite{Leggett}.

In the present study, we examine the metastability of a
current-carrying state with one unit of circulation, or
equivalently, the energetic stability of a vortex state that is
located at the center of the cloud. Our results show that in a
two-dimensional trap, at least in the Thomas-Fermi limit of
strong interactions, the vortex state, viewed as a particle,
feels a force that is proportional to the gradient $-\nabla
n_0(\rho)$, where $n_0(\rho)$ is the single-particle density
distribution of the non-rotating cloud. Here $\rho$ is the
usual radial variable in cylindrical polar coordinates.
Therefore, we conclude that a necessary condition for
metastability is that $n_0(\rho)$ is not a monotonically
decreasing function. We give an analytic derivation of this
result in Sec.\,VI.

For weak interactions metastability is not possible. The
dispersion relation, i.e., the total energy of the cloud ${\cal
E}$ for a given value of the angular momentum per atom $l
\hbar$, is dominated by the single-particle energy, and
therefore local minima cannot appear in ${\cal E}(l)$. In the
Thomas-Fermi limit of strong and repulsive interactions,
metastability is possible, since the interaction energy is much
larger than the single-particle energy. In this limit, the
density $n_0(\rho)$ of the non-rotating cloud is a mirror image
of the trapping potential $V(\rho)$. Therefore, we conclude
that in the Thomas-Fermi limit a necessary condition for
metastability is also that $V(\rho)$ is not a monotonically
increasing function of $\rho$.

In our discussion so far we have distinguished between
$n_0(\rho)$ and $V(\rho)$. However, since metastability
requires that the interaction energy is at least comparable to
the single-particle energy, and in this case a
monotonically-increasing $V(\rho)$ implies a
monotonically-decreasing $n_0(\rho)$, a necessary condition for
metastability is also that $V(\rho)$ is not a monotonically
increasing function of $\rho$.

In what follows we first present our model and our method in
Sec.\,II. Section III describes the results we have derived
within the mean-field approximation, and Sec.\,IV the results
of numerical diagonalization. Section V examines the case of
attractive interactions. In Sec.\,VI we give some analytic
arguments which support our numerical results. Finally, in
Sec.\,VII we consider a driven gas, and conclude that
independently of the form of the trapping potential, any single
off-center vortex state is unstable, while the gas shows
hysteresis.

\section{Model and method}

In general one has to solve a three-dimensional problem.
However, the motion along the axis of rotation introduces an
extra degree of freedom, which may give rise to vortex bending.
To get rid of this complication, we thus consider a
highly-oblate trap, with a very tight, harmonic confinement
along the axis of rotation -- taken to be the $z$ axis -- and
some axially-symmetric trapping potential $V(\rho)$
perpendicular to the axis of rotation,
\begin{equation}
   V_{\rm ext}({\bf r}) = V(\rho) + \frac 1 2 M \omega_z^2 z^2.
\end{equation}
We assume that the corresponding quantum of energy $\hbar
\omega_z$ for motion along the $z$ axis is much larger than any
other energy scale in the problem. The motion of the atoms is
thus quasi-two-dimensional, since their motion along the $z$
axis is frozen.

The energy of the gas at $l=1$, ${\cal E}(l=1)$ is higher than
${\cal E}(l=0)$. The interesting question is whether there is a
barrier in the dispersion relation ${\cal E}(l)$ that separates
the state with $l = 1$ from the state with $l = 0$. In the
presence of such a barrier, the current-carrying state is
metastable, and relatively weak perturbations that do not
conserve the energy and the angular momentum cannot destabilize
it.

We mentioned above that our study examines the energetic
stability of a vortex state that is located at the center of
the cloud. To understand the connection between this kind of
stability criterion, and the calculation of ${\cal E}(l)$, one
should remember that there is an one-to-one correspondence
between the angular momentum per atom $l \hbar$, and the
distance $R_v$ of the vortex state from the center of the trap.
An expression for $l(R_v)$ is derived in Sec.\,VI. The function
${\cal E}(l)$ may thus be viewed also as a function ${\cal
E}(R_v)$, and since $l(R_v=0) = 1$, a local minimum in ${\cal
E}(l)$ for $l \approx 1$, implies that the vortex state is
energetically stable under (at least) infinitesimal
displacements of the vortex from the center of the trap.

To calculate the dispersion relation, we use both the
mean-field Gross-Pitaevskii approximation, as well as numerical
diagonalization of the Hamiltonian. The calculated energies
from the two methods agree to leading order in $N$ \cite{JKMR},
i.e., they differ due to finite-size corrections in the small
systems that we consider in numerical diagonalization.

\section{Mean-field approximation}

Starting with the mean-field, Gross-Pitaevskii approximation,
the form of the trapping potential (which is tight along the
$z$ axis) allows us to use the following ansatz for the order
parameter,
\begin{equation}
 \Psi(x,y,z) = \Phi(x,y) \, \phi_0(z),
\end{equation}
where $\phi_0(z)=e^{-z^2/(2 a_z^2)}/(\pi a_z^2)^{1/4}$ is the
ground state of the harmonic oscillator along the $z$ axis,
with $a_z$ being the oscillator length along this axis. The
corresponding Gross-Pitaevskii equation for the two-dimensional
order parameter $\Phi(x,y)$ is
\begin{eqnarray}
   - \frac {\hbar^2} {2M} (\nabla_x^2 + \nabla_y^2)
 \Phi + V(x,y) \Phi + \tilde g |\Phi|^2 \Phi = \mu \Phi,
\label{en22}
\end{eqnarray}
where $\mu$ is the chemical potential, and $\tilde g = N U_0
\int |\phi_0(z)|^4 dz = \sqrt{8 \pi} \, \hbar^2 N a/M a_z$.
Here $N$ is the number of atoms, and $U_0 = 4 \pi \hbar^2 a/M$,
where $a$ is the s-wave scattering length for elastic atom-atom
collisions, and $M$ is the atom mass. We define for convenience
the dimensionless quantity $g = M \tilde g/ \hbar^2 = \sqrt{8
\pi} N a / a_z$. This parameter may easily get at least as
large as $10^3$, if, for example, $N=10^5$, $a=50$ \AA, and
$a_z \sim 10$ $\mu$m.

The functional whose minimum value we are looking for, is
\begin{eqnarray}
  {\cal E}(\Phi, \Phi^*) =
    - \frac {\hbar^2} {2M} \int \Phi^* (\nabla_x^2 + \nabla_y^2)
   \Phi \, dx dy + \nonumber \\
\int \Phi^* V(x,y)  \Phi \, dx dy
    + \frac {\tilde g} 2 \int |\Phi|^4 \, dx dy,
\label{en221}
\end{eqnarray}
under the extra constraint of a fixed normalization $\int
|\Phi|^2 \, dx dy = 1$. Typically one minimizes the above
expression at a fixed rotational frequency of the trap
$\Omega$, by introducing the new function ${\cal E}_{\rm rot} =
{\cal E} - l \Omega$, where $l = i \int \Phi^* (y \partial_x -
x \partial_y ) \Phi \, dx dy$. If the second derivative of the
(minimized) function $\cal E$ with respect to $l$ is positive,
then one may get the minimized value of ${\cal E} (\Omega)$, as
well as $l(\Omega)$, at fixed $\Omega$, and finally derive the
dispersion relation ${\cal E}(l)$.

However, if the second derivative of the (minimized) function
$\cal E$ with respect to $l$ is negative, within a given range
of $l$, $l_1 \le l \le l_2$, the only minima occur at the end
points $l = l_1$ and $l = l_2$. To overcome this difficulty,
one may minimize the following expression instead \cite{Pap},
\begin{eqnarray}
  E(\Phi^*, \Phi) = {\cal E}(\Phi^*, \Phi)
   + \frac C 2 \hbar \omega (l - l_0)^2,
\label{en222}
\end{eqnarray}
again under the constraint of a fixed normalization $\int
\Phi^* \Phi \, dx dy = 1$. Here $C$ and $l_0$ are real,
dimensionless, and positive constants. The constant $C$ has to
be chosen sufficiently large to ensure that the last term in
Eq.\,(\ref{en222}) is the dominant one, which then also gives
the corresponding minimized function $E(l)$ a positive
curvature, namely ${\cal E}''(l) + C \hbar \omega > 0$, for all
values of the angular momentum $l_1 \le l \le l_2$, where
${\cal E}''(l)$ denotes the second derivative with respect to
$l$. Under the condition of a positive curvature of $E(l)$, one
may minimize $E(l)$ at any $l_{\rm min}$ that satisfies the
equation
\begin{eqnarray}
  l_{\rm min}  = l_0 - {\cal E}'(l_{\rm min})/(C \hbar \omega).
\end{eqnarray}

Using the method of imaginary-time propagation \cite{relax}, we
minimize the expression of Eq.\,(\ref{en222}) for three
different trapping potentials, namely harmonic, anharmonic, and
``Mexican-hat-like", that are plotted in Fig.\,1. In Fig.\,2 we
consider the harmonic trapping potential of the form
\begin{eqnarray}
   V(\rho) = \frac 1 2 M \omega^2 \rho^2,
\label{harm}
\end{eqnarray}
and plot the dispersion relation ${\cal E}(l)$, in units of
$\hbar \omega$, for four values of the parameter $g = 10, 100,
500$, and 1000. In this figure we have shifted the zero of the
energy, so that for all curves ${\cal E}(l=0) = 0$. The actual
shifts of the energies are: for $g=10$, ${\cal E}(l=0) \approx
1.59 \hbar \omega$, for $g=100$, ${\cal E}(l=0) \approx 3.95
\hbar \omega$, for $g=500$, ${\cal E}(l=0) \approx 8.51 \hbar
\omega$ and for $g=1000$, ${\cal E}(l=0) = 11.97 \hbar \omega$.

As one can see in Fig.\,2, the curvature of ${\cal E}(l)$ is
negative for all values of $g$. Although the slope of this
curve at $l = 1^-$ decreases with increasing $g$, our
calculations show that it never becomes zero, and no metastable
minimum forms around $l=1$. Therefore, independently of the
value of the coupling, a harmonic trapping potential cannot
support persistent currents.

To examine the effect of the functional form of $V(\rho)$, we
also consider a more flat potential around the origin, which
has the form
\begin{eqnarray}
    V = \frac 1 2 \hbar \omega (\rho/a_{\perp})^\alpha,
\label{anh}
\end{eqnarray}
where $a_{\perp} = (\hbar/M \omega)^{1/2}$ is the oscillator
length along the plane of motion of the atoms, for $\alpha =
32$, and $g = 30, 100$, and $1000$. Actually, this high value
of $\alpha$ corresponds to a trap which is almost like a
hard-wall potential, with a radius $a_{\perp}$.

Again, we plot ${\cal E}(l)$ in Fig.\,3 shifting the zero of
the energy in each case, so that ${\cal E}(l=0) = 0$. These
shifts are: for $g=30$, ${\cal E}(l=0) \approx 7.41 \hbar
\omega$, for $g=100$, ${\cal E}(L=0) \approx 67.15 \hbar
\omega$, and for $g=1000$, ${\cal E}(L=0) = 124.72 \hbar
\omega$.

This almost hard-wall potential does not allow the gas to
expand radially, as in the case of harmonic confinement. 
The slope of the dispersion relation ${\cal E}(l)$ for $l \to 0$ 
increases with increasing $g$; in contrast for a harmonic 
trapping potential, the slope decreases. This observation 
probably implies that there is a critical power-law dependence 
of the trapping potential, for which this slope does not depend 
on $g$. Also, the slope of ${\cal E}(l)$ for $l \to 1^-$ tends 
to zero for high values of $g$, due to the almost homogeneous 
density of the cloud close to the center of the trap. Apart from 
these observations, ${\cal E}(l)$ has the same qualitative 
features as in the case of harmonic trapping, with no metastable 
minimum forming around $l=1$. Persistent currents are not 
possible in this case, either. More generally, our calculations 
imply that this is also true even for a hard-wall potential, 
$\alpha \to \infty$.

\begin{figure}[t]
\includegraphics[width=8.5cm,height=6cm]{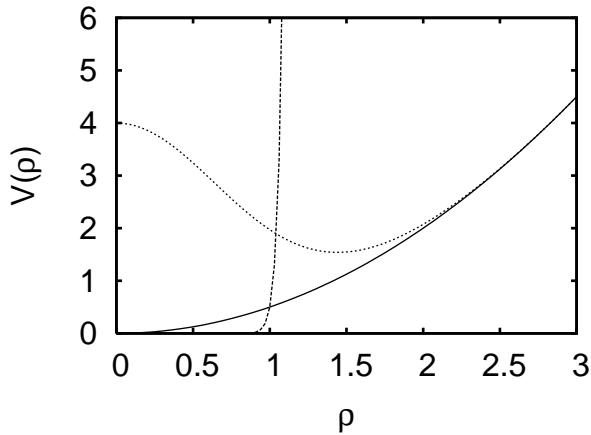}
\caption[]{The three forms of the trapping potential $V(\rho)$
that we use in the present study, namely harmonic,
Eq.\,(\ref{harm}), anharmonic, Eq.\,(\ref{anh}), and ``Mexican
hat", Eq.\,(\ref{mh}). The energy is measured in units of
$\hbar \omega$, and the length in units of $a_{\perp}$.}
\label{FIG1}
\end{figure}

\begin{figure}[t]
\includegraphics[width=8.5cm,height=6cm]{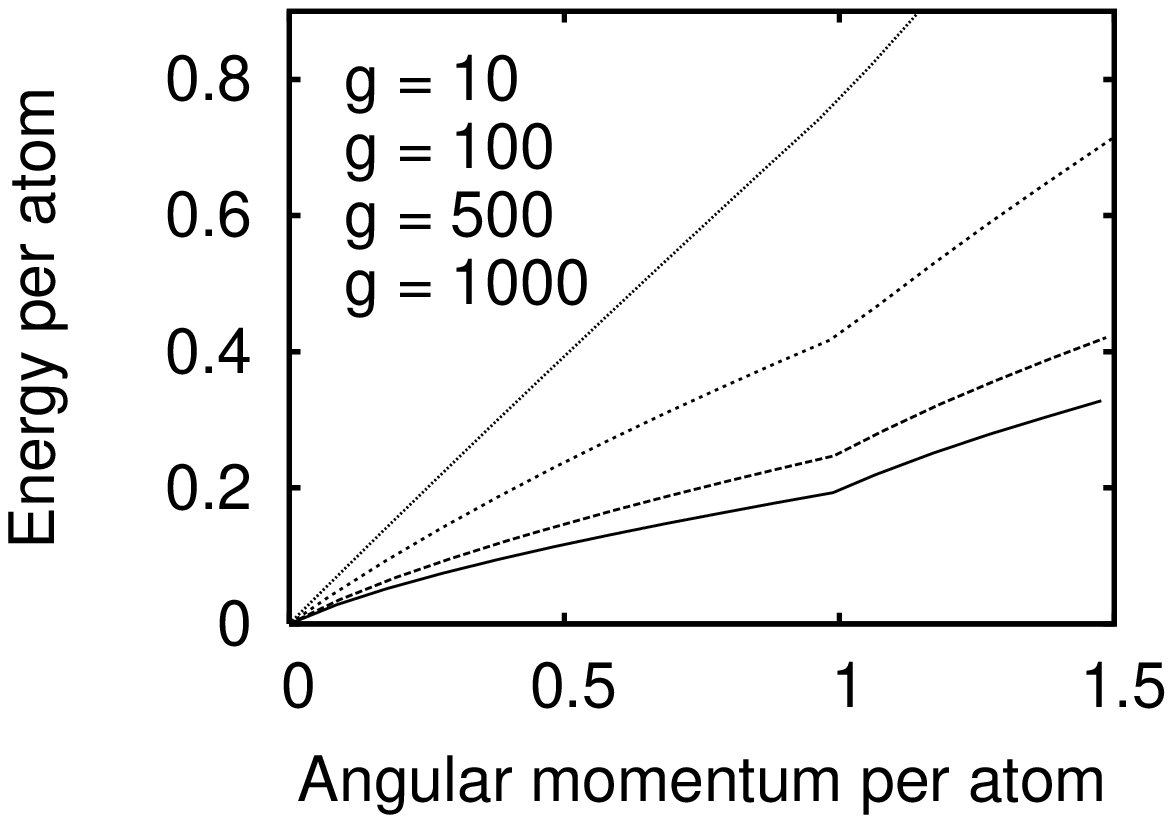}
\includegraphics[width=8.5cm,height=6cm]{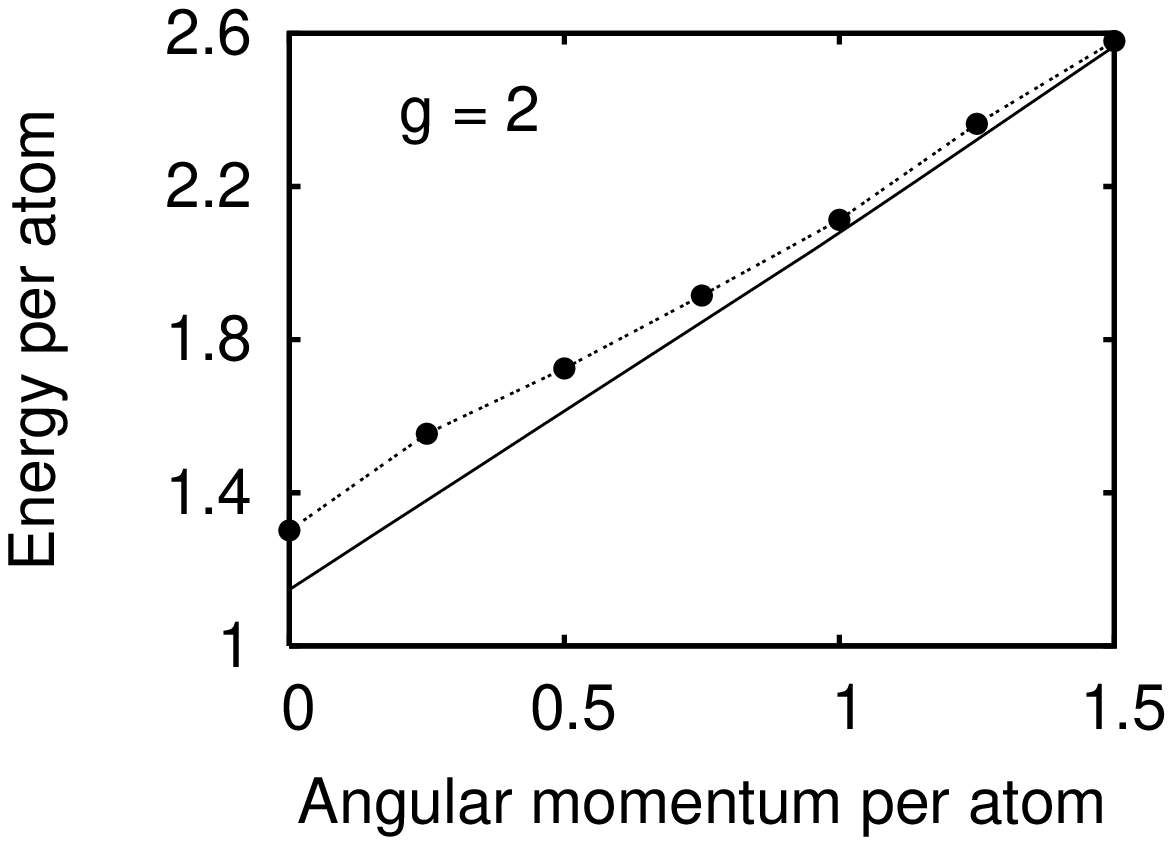}
\caption[]{Top: The dispersion relation ${\cal E}(l)$ in units
of $\hbar \omega$, for a harmonic potential, and for $g=10,
100, 500$, and 1000, calculated within the mean-field
approximation. Bottom: ${\cal E}(l)$ that results from
numerical diagonalization of the Hamiltonian (dots), for $N=4$
atoms and $0 \le L \le 6$ units of angular momentum, as
compared to the one calculated within mean field, for $g=2$.}
\label{FIG2}
\end{figure}

\begin{figure}[t]
\includegraphics[width=8.5cm,height=6cm]{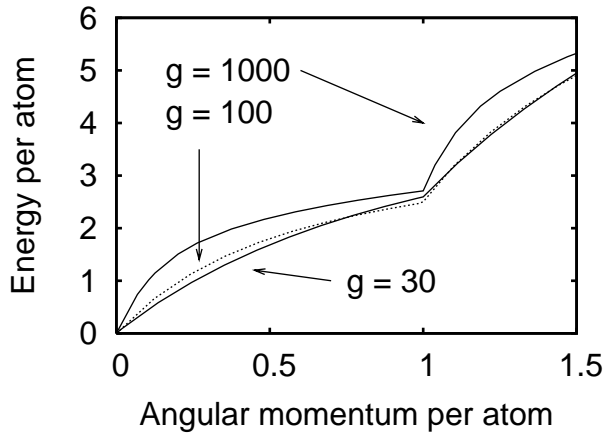}
\caption[]{The dispersion relation ${\cal E}(l)$ in units of
$\hbar \omega$, for an anharmonic potential of the form of
Eq.\,(\ref{anh}), with $\alpha = 32$, and for $g=30, 100$, and
1000, calculated within the mean-field approximation.}
\label{FIG3}
\end{figure}

We turn now to another functional form of $V(\rho)$, which does
not increase monotonically with $\rho$, i.e., we consider a
``Mexican-hat" shape,
\begin{eqnarray}
   V(\rho) = \frac 1 2 M \omega^2 \rho^2 + V_0 e^{-\rho^2/a_\perp^2},
\label{mh}
\end{eqnarray}
with $V_0 > \hbar \omega/2$. Such a potential can be realized
experimentally if one applies a Gaussian, narrow laser beam on
top of the ordinary harmonic trap \cite{Dal}.

Figure 4 shows the corresponding dispersion relation, again
after shifting the zero of the energy at $l=0$, for $V_0/\hbar
\omega = 4$, and $g=2$, 20 and 100. The shifts are: for $g=2$,
${\cal E}(l=0) \approx 2.44 \hbar \omega$, for $g=20$, ${\cal
E}(l=0) \approx 2.94 \hbar \omega$, and for $g=100$, ${\cal
E}(l=0) = 4.50 \hbar \omega$. In this case ${\cal E}(l)$ does
develop a local minimum around $l=1$ for sufficiently strong
interactions, $g \approx 20$. Persistent currents are possible
in this case. In Ref.\,\cite{KYOR} we have shown that in a
strictly one-dimensional trap with periodic boundary
conditions, a similar picture emerges for ${\cal E}(l)$ as the
strength of the interaction increases. This is not a surprising
result, since, as shown in Fig.\,2, the trapping potential
is -- roughly speaking -- toroidal-like in this case.

\begin{figure}[t]
\includegraphics[width=8.5cm,height=6.cm]{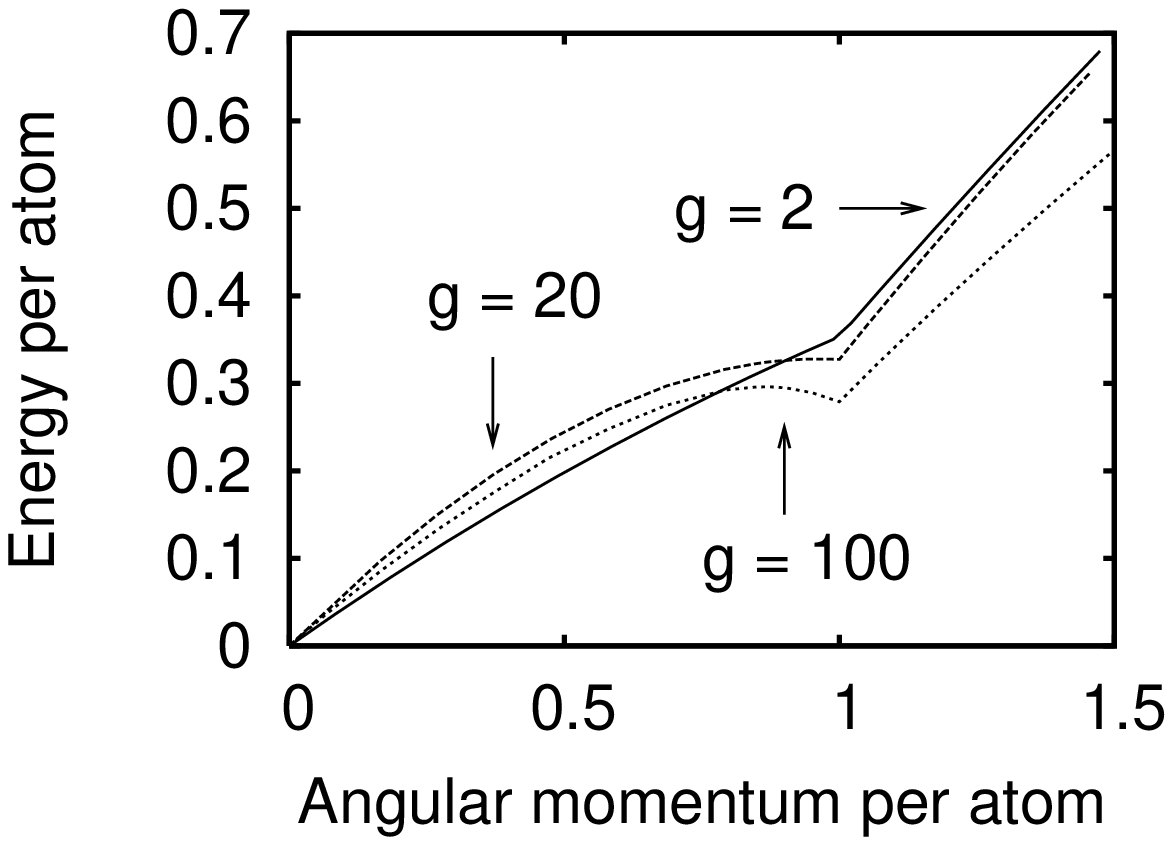}
\includegraphics[width=8.5cm,height=6cm]{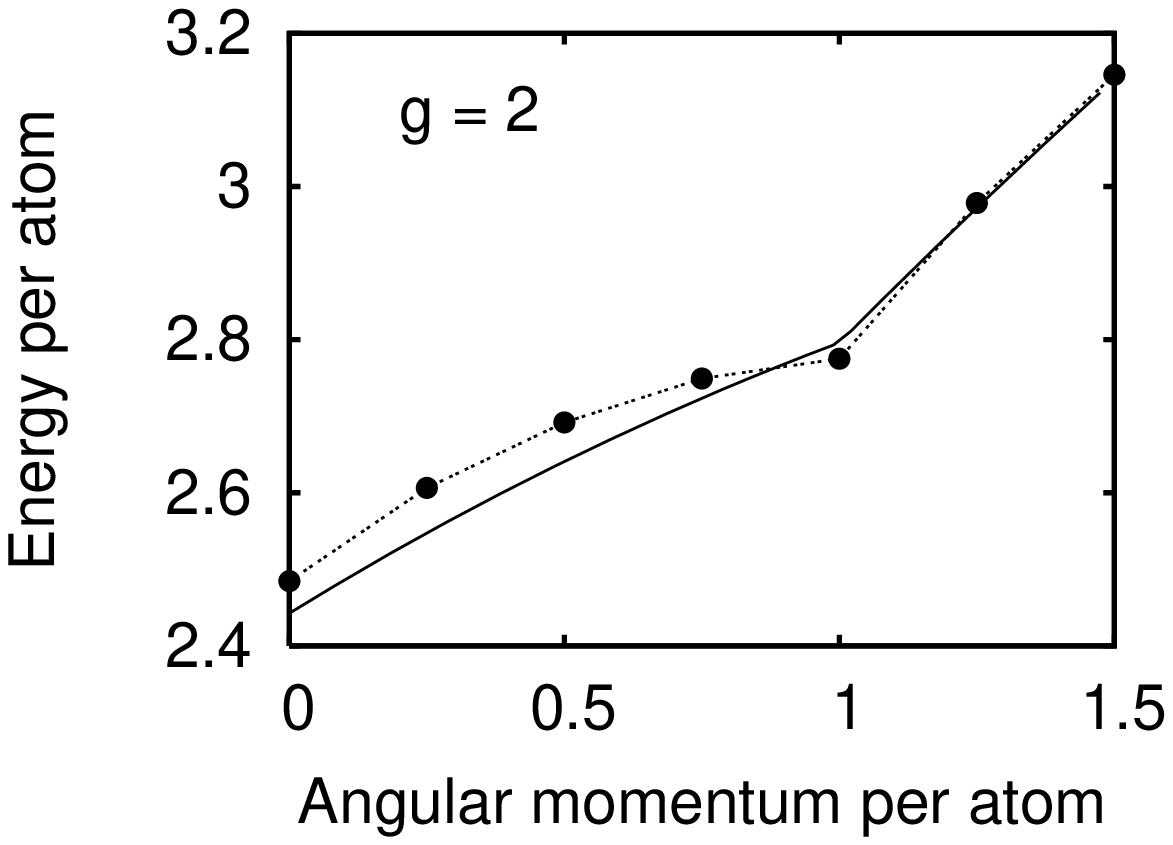}
\caption[]{Top: The dispersion relation ${\cal E}(l)$ in units
of $\hbar \omega$, for a ``Mexican-hat" potential of the form
of Eq.\,(\ref{mh}), with $V_0/\hbar \omega = 4$, and for $g=2,
20$, and 100, calculated within the mean-field approximation.
Bottom: ${\cal E}(l)$ that results from numerical
diagonalization of the Hamiltonian (dots), for $N=4$ atoms and
$0 \le L \le 6$ units of angular momentum, as compared to the
one calculated within mean field, for $g=2$.}
\label{FIG4}
\end{figure}

\section{Numerical diagonalization}

We have also diagonalized numerically the many-body
Hamiltonian, for small systems, of $N$ atoms and $L$ units of
angular momentum. Again we assume tight confinement along the
axis of rotation and essentially consider two-dimensional
motion. We use the eigenstates of the harmonic oscillator in
two dimensions to build the Fock states, which are eigenstates
of the operators $\hat N$ and $\hat L$, and then diagonalize
the resulting Hamiltonian matrix.

Clearly one has to truncate, since the dimensionality of the
matrix becomes very large if one considers highly-excited
states. In the present calculation for each $L$ and $N$, we
include in the Hilbert space as many single-particle states as
are necessary, so that the lowest eigenvalue of the Hamiltonian
has converged up to the third decimal point.

The calculated energies that we obtain agree with those of the
mean-field approximation, apart from finite-size corrections
\cite{JKMR}. The dots in the lower plots of Figs.\,2, and 4
show the corresponding dispersion relation (i.e., the lowest
eigenenergy divided by $N$), as function of the angular
momentum per atom $L/N$, for $0 \le L \le 4$, $N = 4$, and $g =
2$.

\section{Absence of metastability for attractive interactions}

Another interesting question is the possibility of metastable
currents in the case of an effective attractive interaction
between the atoms. Our calculations show that metastability is
not possible in this case, in agreement with the arguments of
Leggett \cite{Leggett}.

In two interesting studies, Wilkin, Gunn, and Smith \cite{WGS},
as well as Mottelson \cite{Ben} have shown that in a harmonic
trap, for an effective interaction that is very weak and
attractive, within the lowest Landau level approximation, the
angular momentum is carried by the center of mass motion
\cite{PP}. The reason is that in a harmonic trap, the center of
mass and relative motions decouple, and as a result the least
energetically-expensive way to give angular momentum to the
gas, is via excitation of the center of mass. The relative
distance between the atoms is thus unaffected, and the
interaction energy does not depend on the angular momentum.

More generally, when the effective interaction is attractive,
but not infinitesimally weak, the lowest, non-rotating state is
already an interesting and non-trivial question. The absolute
minimum of the energy of the gas for negative $g$ corresponds
always to a collapsed state. However, before the cloud
collapses, for sufficiently small values of the parameter
$|g|$, the cloud may be in a metastable state. If the trapping
potential $V(\rho)$ increases monotonically with $\rho$, the
single-particle density of the (non-rotating) cloud is
axially-symmetric, but it shrinks towards the origin of the
trap, because of the attractive interaction. If the trapping
potential does not increase monotonically, the gas may have a
homogeneous density along the axially symmetric minimum of the
potential, or it may form a symmetry-breaking localized blob.

Depending on the lowest state in the absence of rotation, the
current-carrying states may involve vortex excitation, or
center of mass excitation. Our results indicate that,
independently of the form of the trapping potential, and of the
single-particle density distribution, the dispersion relation
is a smooth function of $l$, with a non-negative curvature.

Figure 5 shows a specific example, where we calculate ${\cal
E}(l)$ for a trapping potential that has a ``Mexican-hat"
shape, with $V_0/\hbar \omega = 4$. The density is axially
symmetric around the minimum of $V(\rho)$ for $g=-2$, and it is
localized for $g=-3$, having formed a soliton-like blob, as in
the case of the one-dimensional problem \cite{Ueda,GMK}. As $g$
becomes even smaller, the cloud collapses. In both cases
$g=-2$, and $g=-3$, the dispersion relation has a positive
curvature. In particular when $g=-3$, ${\cal E}(l) \propto
l^2$, since the motion corresponds to solid-body rotation.

\begin{figure}[t]
\includegraphics[width=8.5cm,height=6cm]{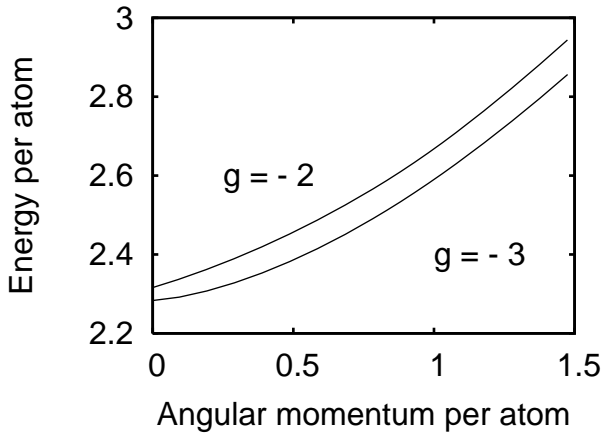}
\caption[]{The dispersion relation ${\cal E}(l)$ in units of
$\hbar \omega$, calculated within the mean-field approximation,
for a ``Mexican-hat" potential of the form of Eq.\,(\ref{mh}),
with $V_0/\hbar \omega = 4$, and for $g=-2$ (higher curve), and
$g = -3$ (lower curve).}
\label{FIG5}
\end{figure}

\section{Analytical arguments}

The results of Secs.\,III, IV and V show that metastability of
superflow is possible for sufficiently strong and repulsive
interactions. Furthermore, when the single-particle density
$n_0(\rho)$ of the non-rotating cloud decreases monotonically,
${\cal E}(l)$ is a monotonically-increasing function and local
minima in the dispersion relation are not possible. We give
here a simple derivation of this result.

The dispersion relation ${\cal E}(l)$ consists of the
single-particle energy and of the interaction energy, ${\cal E}
= \langle H_{\rm sp} \rangle + \langle V_{\rm int} \rangle$.
The contribution of the single-particle part of the Hamiltonian
$H_{\rm sp}$ is a smooth, and increasing function of $l$, with
$\partial \langle H_{\rm sp} \rangle / \partial l > 0$. We thus
focus on the interaction energy $\langle V_{\rm int} \rangle$,
i.e., on the quantity $\langle V_{\rm int} \rangle = (g/2) \int
|\Phi|^4 \, dx dy$.

We consider the derivative
\begin{eqnarray}
  \frac {\partial \langle V_{\rm int} \rangle}{\partial l}
  = \tilde{g} \int |\Phi|^2
  \frac {\partial |\Phi|^2} {\partial l} \, dx dy.
\label{integral}
\end{eqnarray}
Clearly, the sign of the above derivative is determined by the
sign of $\partial |\Phi|^2/\partial l$. The density that enters
in the above expression is the local density of the cloud
$n(\rho,\phi)=|\Phi|^2$. In the limit of strong interactions,
$g \gg 1$, the coherence length $\xi$ is much smaller than the
radius of the cloud $R_0$, with $\xi / R_0 \sim g^{-1/2}$. The
distortion of the cloud (due to the vortex) is localized around
a region of radius $\approx \xi$. For this reason, in the
calculation of the integral in Eq.\,(\ref{integral}) one may
use the density profile neglecting the density depression due
to the vortex, and also the weak dependence of the density of
the gas on the polar angle $\phi$. Denoting the corresponding
density as $n(\rho)$ [which is qualitatively the same as
$n_0(\rho)$],
\begin{eqnarray}
  \frac {\partial \langle V_{\rm int} \rangle}{\partial l}
  \approx \tilde{g} \int n(\rho)
  \frac {\partial n(\rho)} {\partial l} \, dx dy.
\label{integral2}
\end{eqnarray}
Then, using the chain rule
\begin{eqnarray}
  \frac {\partial n} {\partial l} = ({\hat \rho} \cdot
  \nabla{n})
  \frac {\partial R_v} {\partial l},
\label{deriv}
\end{eqnarray}
where $R_v$ is the distance of the vortex from the center of
the cloud, we observe that when $n(\rho)$ decreases
monotonically, the inner product ${\hat \rho} \cdot
\nabla{n_0}$ is always negative.

Turning to the derivative ${\partial R_v} /{\partial l}$, this
is negative. The function $R_v(l)$ is monotonically-decreasing,
with $R_v=0$ for $l=1$, and $R_v = \infty$, when $l=0$.
Generalizing the result of Ref.\,\cite{PS} that is given for a
homogeneous system, we derive an approximate expression for
$l(R_v)$. More specifically the angular momentum of the gas
around the center of the trap is (in units of $\hbar$),
\begin{equation}
  l = \frac 1 \hbar \int M \rho v \, n(\rho) \, \rho d \rho d \phi.
\label{llvr}
\end{equation}
However, the circulation
\begin{equation}
  \oint {\bf v} \cdot d{\bf w} = \frac h M,
\end{equation}
provided that the vortex is inside the area that is defined by
the corresponding line $\bf w$. Considering a circle of radius
$\rho$,
\begin{equation}
   \int v \rho d \phi = \frac h M,
\label{circ}
\end{equation}
for $\rho > R_v$ and zero for $\rho < R_v$. From
Eqs.\,(\ref{llvr}) and (\ref{circ}) we get that,
\begin{eqnarray}
  l(R_v) = 2 \pi \int_{R_v}^{\infty} n(\rho) \rho d \rho
   = \int_{R_v}^{\infty} n(\rho) \, d^2 \rho,
\label{lr}
\end{eqnarray}
which is the fraction of the atoms that reside outside a circle
of radius $R_v$. Equation (\ref{lr}) implies that $l(R_v)$ is a
decreasing function, and therefore $\partial R_v/\partial l <
0$. An analytic expression for $l(R_v)$ may be derived for weak
interactions (i.e., within the lowest-Landau level
approximation), which is $l(R_v) \approx 1 - R_v^2/ (2
a_{\perp}^2)$ \cite{KMP}, when the vortex state is close to the
center of the trap (for small values of $R_v$).

We conclude from Eqs.\,(\ref{integral2}), (\ref{deriv}) and
(\ref{lr}), that when $n_0(\rho)$ decreases monotonically,
$\partial \langle V_{\rm int} \rangle/\partial l > 0$, and
thus
\begin{eqnarray}
  \frac {\partial {\cal E}} {\partial l} = \frac {\partial \langle
H_{\rm sp} \rangle} {\partial l} + \frac {\partial \langle
V_{\rm int} \rangle} {\partial l} > 0,
\end{eqnarray}
which means that indeed metastability is not possible in this
case.

As mentioned also earlier, in the Thomas-Fermi limit one may go
even further with this argument, since $n_0(\rho)$ is roughly
the mirror image of the trapping potential $V(\rho)$.
Therefore, in the Thomas-Fermi limit, a necessary condition for
metastability of superflow is that $V(\rho)$ does not increase
monotonically with $\rho$. This condition holds even for more
weak interactions, i.e., in the cross-over region, when the
single-particle energy is comparable to the interaction energy.

\section{The behavior of a driven gas and some general conclusions}

In addition to the question of metastability of superflow, the
calculated ${\cal E}(l)$ for the various forms of $V(\rho)$
also allow us to make some general statements for the behavior
of a driven gas. Remarkably these statements are independent of
$V(\rho)$.

More specifically, one may define three critical rotational
frequencies, namely the slope $\Omega_1$ of ${\cal E}(l)$ at $l
= 1^-$, the frequency $\Omega_2$ at which $\hbar \Omega_2 =
{\cal E}(l=1) - {\cal E}(l=0)$, and the slope $\Omega_3$ of
${\cal E}(l)$ at $l=0$. According to our calculations, in all
cases, $\Omega_1 < \Omega_2 < \Omega_3$. In a harmonic, and
highly-oblate trap, it has been shown that in the Thomas-Fermi
limit, $\Omega_2 = (5/3) \Omega_1$ \cite{Fetter}. As we explain
below, the fact that $\Omega_1 < \Omega_2 < \Omega_3$ has some
important implications.

Experimentally one may either cool down the gas to very low
temperatures and then rotate, or first rotate above the
condensation temperature, and then cool down. In the first
case, as one starts rotating the trap (at a fixed $g$), the
cloud undergoes a discontinuous transition from the
non-rotating state to the state with one vortex that is located
at the center of the trap, at the frequency $\Omega =
\Omega_3$. In the reverse process, the vortex leaves the cloud
discontinuously at an $\Omega = \Omega_1$, and therefore, there
is hysteresis. In addition, for $\Omega_1 < \Omega < \Omega_2$,
the vortex state is metastable, i.e., it becomes locally stable
before it becomes thermodynamically stable (at $\Omega =
\Omega_2$). Finally, because of the negative curvature of
${\cal E}(l)$, any single off-center vortex state is unstable.

In the second case, if one first rotates and then cools down, a
(meta)stable vortex state will appear at the center of the
cloud at a rotational frequency $\Omega = \Omega_1$. Again, any
single off-center vortex state is unstable.

The physical picture that underlies the above arguments for the
metastability, the hysteresis, and the fact that any off-center
vortex state is unstable, is that the vortex state creates a
node at the density of the gas. As long as the density
$n_0(\rho)$ decreases monotonically, this node cannot produce
any energy barrier \cite{Rokhsar,Leggett}. On the other hand,
if the density is not a monotonically-decreasing function of
$\rho$, there may be an energy barrier in the dispersion
relation, provided that the interaction is repulsive and
sufficiently strong.

\acknowledgements

We thank M. Mageiropoulos for useful discussions. We acknowledge 
financial support from the European Community project ULTRA-1D 
(NMP4-CT-2003-505457), the Swedish Research Council, the Swedish 
Foundation for Strategic Research, and the NordForsk Nordic Network 
on ``Low-dimensional physics".

\end{document}